\documentclass[10pt]{article}
\usepackage{epsfig}
\begin{document}
\newcommand{\noi}{\noindent}
\newtheorem{opgave}{Excercise}
\newcommand{\bop}[1]{\begin{opgave}\begin{rm}{\bf#1} \newline\noi}
\newcommand{\eop}{\end{rm}\end{opgave}}

\newcommand{\nc}{\newcommand}
\nc{\dia}[1]{\mbox{{\Large{\bf diag}}}[#1]}
\nc{\diag}[3]{\raisebox{#3cm}{\epsfig{figure=#1.eps,width=#2cm}}}
%\nc{\diag}[3]{\dia{#1}}
\nc{\bq}{\begin{equation}}
\nc{\eq}{\end{equation}}
\nc{\bqa}{\begin{eqnarray}} 
\nc{\eqa}{\end{eqnarray}}
\nc{\eqan}{\nonumber\end{eqnarray}}
\nc{\nl}{\nonumber \\}
\nc{\grf}{Green's function}
\nc{\grfs}{Green's functions}
\nc{\cgrf}{connected Green's function}
\nc{\cgrfs}{connected Green's functions}
\nc{\f}{\varphi}
\nc{\exb}[1]{\exp\!\left(#1\right)}
\nc{\avg}[1]{\left\langle #1\right\rangle}
\nc{\suml}{\sum\limits}
\nc{\prodl}{\prod\limits}
\nc{\intl}{\int\limits}
\nc{\ddv}[1]{{\partial\over\partial #1}}
\nc{\ddvv}[2]{{\partial^{#1}\over\left(\partial #2\right)^{#1}}}
\nc{\la}{\lambda}
\nc{\La}{\Lambda}
\nc{\eqn}[1]{Eq.(\ref{#1})}
\nc{\eqns}[1]{Eqs.(\ref{#1})}
%\nc{\appndix}[3]{\section{Appendix #1 : #2\label{#3}}\input{#3}}
\nc{\appndix}[3]{\section{{#2}\label{#3}}\input{#3}}
\nc{\cala}{{\cal A}}
\nc{\calb}{{\cal B}}
\nc{\cale}{{\cal E}}
\nc{\calf}{{\cal F}}
\nc{\calh}{{\cal H}}
\nc{\calc}{{\cal C}}
\nc{\cald}{{\cal D}}
\nc{\calp}{{\cal P}}
%\nc{\calphi}{{\cal{\Phi}}}
\nc{\calphi}{\Psi}
\nc{\lra}{\;\leftrightarrow\;}
\nc{\stapel}[1]{\begin{tabular}{c} #1\end{tabular}}
\nc{\al}{\alpha}
\nc{\be}{\beta}
\nc{\g}{\gamma}
\nc{\ga}{\gamma}
\nc{\ka}{\kappa}
\nc{\om}{\omega}
\nc{\si}{\sigma}
\nc{\Si}{\Sigma}
\nc{\ro}{\rho}
\nc{\vph}{\vphantom{{A^A\over A_A}}}
\nc{\wph}{\vphantom{{A^A}}}
\nc{\order}[1]{{\cal O}\left(#1\right)}
\nc{\De}{\Delta}
\nc{\de}{\delta}
\nc{\vx}{\vec{x}}
\nc{\vxi}{\vec{\xi}}
\nc{\vy}{\vec{y}}
\nc{\vk}{\vec{k}}
\nc{\vq}{\vec{q}}
\nc{\vp}{\vec{p}}
\nc{\lijst}[1]{\begin{center}
  \fbox{\begin{minipage}{12cm}{#1}\end{minipage}}\end{center}}
\nc{\lijstx}[2]{\begin{center}
  \fbox{\begin{minipage}{#1cm}{#2}\end{minipage}}\end{center}}
\nc{\eps}{\epsilon}
\nc{\caput}[2]{\chapter{#1}\input{#2}}
\nc{\pd}{\partial}
\nc{\dtil}{\tilde{d}}
\nc{\vskp}{\vspace*{\baselineskip}}
\nc{\coos}{co\-ord\-in\-at\-es}
\nc{\calm}{{\cal M}}
\nc{\calj}{{\cal J}}
\nc{\call}{{\cal L}}
\nc{\calt}{{\cal T}}
\nc{\calu}{{\cal U}}
\nc{\calw}{{\cal W}}
\nc{\msq}{\langle|\calm|^2\rangle}
\nc{\nsym}{F_{\mbox{{\tiny{symm}}}}}
\nc{\dm}[1]{\mbox{\bf dim}\!\left[#1\right]}
\nc{\fourv}[4]{\left(\begin{tabular}{c}
  $#1$\\$#2$\\$#3$\\$#4$\end{tabular}\right)}
\nc{\fs}[1]{/\!\!\!#1}
\nc{\dbar}[1]{{\overline{#1}}}
\nc{\tr}[1]{\mbox{Tr}\left(#1\right)}
\nc{\row}[4]{$#1$ & $#2$ & $#3$ & $#4$}
\nc{\matv}[4]{\left(\begin{tabular}{cccc}
   #1 \\ #2 \\ #3 \\ #4\end{tabular}\right)}
\nc{\twov}[2]{\left(\begin{tabular}{cccc}
   $#1$ \\ $#2$\end{tabular}\right)}
\nc{\ubar}{\dbar{u}}
\nc{\vbar}{\dbar{v}}
\nc{\lng}{longitudinal}
\nc{\pol}{polarization}
\nc{\longpol}{\lng\ \pol}
\nc{\bnum}{\begin{enumerate}}
\nc{\enum}{\end{enumerate}}
\nc{\nubar}{\overline{\nu}}
\nc{\mau}{{m_{\mbox{{\tiny U}}}}}
\nc{\mad}{{m_{\mbox{{\tiny D}}}}}
\nc{\qu}{{Q_{\mbox{{\tiny U}}}}}
\nc{\qd}{{Q_{\mbox{{\tiny D}}}}}
\nc{\vau}{{v_{\mbox{{\tiny U}}}}}
\nc{\aau}{{a_{\mbox{{\tiny U}}}}}
\nc{\aad}{{a_{\mbox{{\tiny D}}}}}
\nc{\vad}{{v_{\mbox{{\tiny D}}}}}
\nc{\mw}{{m_{\mbox{{\tiny W}}}}}
\nc{\mv}{{m_{\mbox{{\tiny V}}}}}
\nc{\mz}{{m_{\mbox{{\tiny Z}}}}}
\nc{\mh}{{m_{\mbox{{\tiny H}}}}}
\nc{\gw}{{g_{\mbox{{\tiny W}}}}}
\nc{\qw}{{Q_{\mbox{{\tiny W}}}}}
\nc{\qc}{{Q_{\mbox{{\tiny c}}}}}
\nc{\law}{{\Lambda_{\mbox{{\tiny W}}}}}
\nc{\obar}{\overline}
\nc{\guuh}{{g_{\mbox{{\tiny UUH}}}}}
\nc{\gwwz}{{g_{\mbox{{\tiny WWZ}}}}}
\nc{\gwwh}{{g_{\mbox{{\tiny WWH}}}}}
\nc{\gwwhh}{{g_{\mbox{{\tiny WWHH}}}}}
\nc{\gzzh}{{g_{\mbox{{\tiny ZZH}}}}}
\nc{\gvvh}{{g_{\mbox{{\tiny VVH}}}}}
\nc{\ghhh}{{g_{\mbox{{\tiny HHH}}}}}
\nc{\ghhhh}{{g_{\mbox{{\tiny HHHH}}}}}
\nc{\gzzhh}{{g_{\mbox{{\tiny ZZHH}}}}}
\nc{\thw}{\theta_W}
\nc{\sw}{{s_{\mbox{{\tiny W}}}}}
\nc{\cw}{{c_{\mbox{{\tiny W}}}}}
\nc{\ward}[2]{\left.#1\right\rfloor_{#2}}
\nc{\vpa}{(1+\g^5)}
\nc{\vma}{(1-\g^5)}
\nc{\gwzh}{{g_{\mbox{{\tiny WZH}}}}}
\nc{\gudh}{{g_{\mbox{{\tiny UDH}}}}}
\nc{\gwcj}{{g_{\mbox{{\tiny Wcj}}}}}
\nc{\gwcjg}{{g_{\mbox{{\tiny Wcj$\g$}}}}}
\nc{\gwwcc}{{g_{\mbox{{\tiny WWcc}}}}}
\nc{\gzcc}{{g_{\mbox{{\tiny Zcc}}}}}
\nc{\none}[1]{ }
\nc{\Vir}{{V_{\mbox{{\small IR}}}}}
\nc{\kt}[1]{\left|#1\right\rangle}
\nc{\br}[1]{\left\langle#1\right|}
\nc{\tw}{\tilde{w}}
\nc{\hDe}{\nabla}
\nc{\bino}[2]{\left(\begin{tabular}{c}$#1$\\$#2$\end{tabular}\right)}
\nc{\Om}{\Omega}
\nc{\Omb}{\overline{\Omega}}
\nc{\qbar}{\bar{q}}
\nc{\leeg}[1]{}
\nc{\geuler}{\g_{\mbox{{\tiny E}}}}

\begin{center}
{\bf{\Large Ambiguities in Pauli-Villars regularization}}\\
\vspace*{2\baselineskip}
R.H.P. Kleiss\footnote{{\tt R.Kleiss@science.ru.nl}}, T. Janssen\footnote{{\tt thwjanssen89@gmail.com}}\\
Radboud University Nijmegen,\\
Institute for Mathematics, Astrophysics and Particle Physics, \\Heyendaalseweg 135, 
NL-6525 AJ Nijmegen, The Netherlands.\\
\vspace*{3\baselineskip}
{\bf Abstract}\\
\vspace*{\baselineskip}
\begin{minipage}{10cm}{{\small We investigate regularization of scalar one-loop
integrals in the Pauli-Villars subtraction scheme. The results depend on
the number of subtractions, in particular the finite terms that survive after
the divergences have been absorbed by renormalization. Therefore
the process of Pauli-Villars regularization is ambiguous. We discuss
how these ambiguities may be resolved by applying an asymptotically large
number of subtractions, which results in a regularization that is automatically
valid in any number of dimensions.}}\end{minipage}
\vspace*{2\baselineskip}
\end{center}

%\section{Introduction}
The regularization method of Pauli-Villars (PV) subtraction \cite{PV} is of long standing in
quantum field theory. Although not suited to all possible problems (notably, nonabelian gauge theories), 
it is still used in a variety of applications \cite{hiller,brodsky}.
Essentially, PV regularization consists in pairing particle propagators
with (possibly unphysical) propagators of fictitious heavy particles. In \cite{tHV} these
are introduced under the appelation of {\em unitary regulators}. If $m$ is the mass of
the physical (scalar) particle with momentum $q^\mu$, 
and $M$ that of a fictitious heavy particle, PV involves the
modification of the propagator as follows:
\bq
{1\over q^2 - m^2 + i\eps}\;\;\;\to\;\;\;{1\over q^2 - m^2 + i\eps} -{1\over q^2 - M^2 + i\eps}\;\;,
\eq 
thereby reducing the large-$q$ behaviour from $(q^2)^{-1}$ to $(q^2)^{-2}$ and thus improving the
integrability properties of diagrams. At the end of the calculation the limit $M\to\infty$ is
implied.
With this method, the one-loop diagram from a scalar self-interacting theory
\bq\diag{PVdiag1}{1.5}{0}\label{firstdiagram}\eq
becomes integrable in four dimensions. On the other hand, the two diagrams
\bq\diag{PVdiag2}{1}{-.4}\;\;\;\mbox{and}\;\;\;\diag{PVdiag3}{1.7}{-.4}\label{seconddiagrams}\eq
are not integrable in four and six dimensions, respectively. In that case additional
subtractions with a spectrum of fictitious particles are necessary.

It is tempting to perform, for a given diagram, precisely so many PV subtractions as are
necessary to make the loop integral convergent: once for the diagram of \eqn{firstdiagram},
and twice for those of \eqn{seconddiagrams}. But this, of course, runs counter to
the idea of quantum field theory, in which Feynman diagrams themselves have no independent
status but only their combination into Green's functions counts. If there is even only a single
diagram that calls for a double subtraction, say, then {\em all\/} diagrams should
undergo the same double subtraction, even if they are already convergent after a single one.
An approach in which for each diagram precisely sufficient subytactions are made to make
that diagram convergent must be considered  incorrect.\\

This leads to the following question: how do we decide to stop making additional PV
subtractions? A priori there is nothing that forbids one from making very many
subtractions even if that is, strictly speaking, unnecessary.

In what follows, we show that the results of divergent one-loop integrals depend on the number of
PV subtractions, and are therefore ambiguous, 
while loop integrals that are themselves convergent do not. We study
this dependence, and point to a possible resolution of these ambiguities.\\

%\section{Multiple subtractions}
As remarked above, depending on circumstances, a single PV subtraction may not be sufficient to regularize loop
integrals, and multiple subtractions become necessary. 
In what follows, we shall use the abbreviations $\mu=m^2$ and $\La = M^2$.
The $k$-fold PV subtraction of a propagator is defined as follows :
\bqa {1\over s+\mu} &\to& \left\lfloor{1\over s+\mu}\right\rfloor_{PV(k)}\nl &=&
{1\over s+\mu} - {\al_1\over s+\La_1} - {\al_2\over s+\La_2} - \cdots - {\al_k\over s+\La_k} \nl &=& 
{C\over(s+\mu)(s+\La_1)(s+\La_2)\cdots(s+\La_k)}\;\;.\label{firstform}
\eqa
The requirement is that $C$ be independent of $s$. By first considering the limit $s\to-\mu$ 
and then $s\to-\La_r$ (assuming the $\La$'s to be all different) we find immediately that
\bq
C = \prodl_{j=1}^k(\La_j-\mu)\;\;\;,\;\;\;
\al_r = \left(\prodl_{j\ne r}(\La_j-\mu)\right)\left(\prodl_{j\ne r}(\La_j-\La_r)\right)^{-1}\;\;.
\eq
By subtracting sufficiently many PV propagators we can achieve any desired
high-momentum behaviour for the propagator.

That the above subtraction involves $k$  different mass scales is of course lacking in elegance.
We can therefore take $\La_j\to\La$, and then we obtain
\bq
\left\lfloor{1\over s+\mu}\right\rfloor_{PV(k)} = {\De^k\over(s+\mu)(s+\La)^k}\;\;,
\eq
with $\De=\La-\mu$. This can also be written as
\bq
\left\lfloor{1\over s+\mu}\right\rfloor_{PV(k)} = 
{1\over s+\mu} - {1\over s+\La} - {\De\over(s+\La)^2} - \cdots - {\De^{k-1}\over(s+\La)^k}\;\;.
\label{secondform}
\eq
A similar observation was made, for instance, in \cite{stoilov}.
To understand the relation between \eqn{firstform} and \eqn{secondform} it is illustrative to consider the
case $k=2$. Taking $\La_1=\La\;,\;\La_2 = \La+\de$ with small $\de$, we see that
\bqa
\left\lfloor{1\over s+\mu}\right\rfloor_{PV(2)} &=&
{1\over s+\mu} - {\La_2-\mu\over\La_2-\La_1}{1\over s+ \La_1} 
 - {\La_1-\mu\over\La_1-\La_2}{1\over s+ \La_2}\nl
&=& {1\over s+\mu} - {\De+\de\over\de}{1\over s+\La} + {\De\over\de}{1\over s+ \La+\de}\nl
&=& {1\over s+\mu} - {1\over s+\La} - {\De\over(s+\La)^2} + \order{\de}\;\;.
 \eqa
%&=& {1\over s+\mu} - {\De+\de\over\de}{1\over s+\La} + {\De\over\de}
%\left({1\over s+ \La} -\de{1\over(s+\La)^2} + \order{\de^2}\right)\nl
For higher $k$ analogous expansions result. Moreover the result is quite
independent from the precise way in which the limit $\La_j\to\La$ ($j=1,\ldots,n$) is reached.
Note the following fact : if we keep subtracting without limit, we formally have
\bqa
\lim_{k\to\infty}\left\lfloor{1\over s+\mu}\right\rfloor_{PV(k)} &=&
{1\over s+\mu} - {1\over s+\La}\suml_{j=0}^\infty\left({\De\over s+\La}\right)^j\nl
&=& {1\over s+\mu} - {1\over s+\La - \De} = 0\;\;.
\eqa
While, indeed, simply replacing every loop integral by zero makes all loop
corrections trivial, this is clearly not what we want.
Obviously, we must investigate the dependence of loop-integral results on the number
of subtractions.\\

%\section{PV-regularized integrals}
We shall restrict ourselves to one-loop computations in the context of an effective
potential. Therefore no external momenta are involved, and all propagators have the
form $1/(s+\mu)$, where $s=q^2$, $q$ being the loop momentum. 
The (even) 
number of dimensions of spacetime dimensions is denoted by $2\om$. After performing the Wick rotation and 
the angular integral of the loop momentum, the loop integral with $n$ propagators is given by
\bq
J_{\om,n} = \intl_0^\infty ds\;s^{\om-1}\;{1\over(s+\mu)^n}\;\;\;,\;\;\;n>0,
\label{PVJeen}
\eq
where we have dropped any overall factors. This integral is finite if $n>\om$. If this is not the case, 
we must subtract, although strictly speaking we {\em ought\/} to be allowed to PV-subtract
convergent integrals as well. We therefore should replace \eqn{PVJeen} by
\bq
J_{\om,n}^{(k)} = \intl_0^\infty ds\;s^{\om-1}\;\left\lfloor{1\over s+\mu}\right\rfloor_{PV(k)}^n\;\;.
\eq
The integral will be convergent for finite $\La$ if
\bq
k > {\om\over n} - 1\;\;\;\;\;\;\om,n,k\;\;\mbox{integers}\;\;.
\eq
We shall be slightly more general and define the generating function
\bqa
H_{\om,n}(x) &=& \intl_0^\infty ds\;h_{\om,n}(x,s)\;\;,\nl
h_{\om,n}(x,s) &=& {s^{\om-1}\over(s+\mu)^n}\suml_{k\ge1+\om-n}{\De^k\over(s+\La)^k}\,x^k\nl
&=& {s^{\om-1}(x\De)^{1+\om-n}\over(s+\mu)^n(s+\La)^{\om-n}(s+\La - x\De)}\;\;.
\eqa
From the fact that all the functions $h$ decrease as $1/s^2$ for large $s$ we see that we have 
subtracted sufficiently often to make the integrals convergent for finite $\La$ (and hence
PV-regularized). For higher values of $k$, we simply have additional subtractions. The integrals $H$
have series expansions in $x$ : the regularized integral $J_{\om,n}^{(k)}$ is then given as the
coefficient of $x^{nk}$ in $H_{\om,n}(x)$.\\

%\section{Integral ambiguities}
For given $\om$ and $n$ it is a simple matter to integrate $h_{\om,n}(x,s)$ over $s$,
where by construction the upper endpoint $s=\infty$ never contributes. The most 
important point is to note that
\bqa
\log(\La - x\De) %&=& \log(\La) + \log(1-x) + \log\left(1 + {\mu\over\La}{x\over1-x}\right)\nl
&=& \log(\La) + \log(1-x) + {\mu\over\La}{x\over1-x} - {1\over2}\left({\mu\over\La}{x\over1-x}\right)^2
+  \cdots
\eqa
It is already clear that at least some of the $\log(\La)$ terms will be accompanied by $\log(1-x)$.
Below, we give the results for $H_{\om,n}(x)$ for
the most relevant cases for 2,4, and 6 dimensions. By $L$ we
denote $\log(\Lambda/\mu)$. We have taken $\mu/\Lambda$ to zero wherever possible, although this
has of course to be done more carefully in the case of two-loop computations. \bqa
H_{1,1}(x) &=&  {x\over1-x}L + {x\log(1-x)\over1-x}\;\;\;,\nl
H_{1,2}(x) &=& {1\over1-x}\,{1\over\mu}\;\;\;,\nl
H_{2,1}(x) &=& - x\log(1-x)\La - {x^2\over1-x}\mu L - {x^2\over1-x}\left(\log(1-x)+1\wph\right)\mu\;\;\;,\nl
H_{2,2}(x) &=& {x\over1-x}L + {x\over1-x}\left(\wph\log(1-x)-1\right)\;\;\;,\nl
H_{2,3}(x) &=& {1\over1-x}\,{1\over2\mu}\;\;\;,\nl
H_{3,1}(x) &=& \left(\wph x(1-x)\log(1-x)+x^2\right)\La^2 + 2x^2\log(1-x)\,\mu\La\nl && 
+ {x^3\over1-x}\,\mu^2L + {x^3\over1-x}\left(\log(1-x)+{3\over2}\right)\,\mu^2\;\;\;,\nl
H_{3,2}(x) &=& -x\log(1-x)\Lambda - {2x^2\over1-x}\,\mu L - {x(1+x)\log(1-x)\over1-x}\,\mu\;\;\;,\nl
H_{3,3}(x) &=& {x\over1-x}L + {x\over1-x}\left(\log(1-x)-{3\over2}\right)\;\;\;,\nl
H_{3,4}(x) &=& {1\over1-x}\,{1\over3\mu}\;\;\;.
\label{Hvalues}
\eqa
From these we can find the $k$-fold PV-regularized integrals $J^{(k)}_{\om,n}$.
These contain the digamma function $\psi(z)$, defined as
\bq
\psi(z) \equiv {d\over dz}\log\Gamma(z)\;\;\;,\;\;\;
\psi(q) = -\geuler + \suml_{\ell=1}^{q-1}{1\over\ell}\;\;\;,\;\;\;\geuler\approx0.577\ldots
\eq
for integer argument $q$. The various integrals $J$ now read
\bqa
J_{1,1}^{(k\ge1)} &=& L - \psi(k) -\geuler\;\;\;,\nl
J_{1,2}^{(k\ge0)} &=& {1\over\mu}\;\;\;,\nl
J_{2,1}^{(k\ge2)} &=& {\La\over k-1} - \mu L + \mu\left(\wph\psi(k-1) + \geuler -1\right)\;\;\;,\nl
J_{2,2}^{(k\ge1)} &=& L - \psi(2k) -\geuler - 1\;\;\;,\nl
J_{2,3}^{(k\ge0)} &=& {1\over2\mu}\;\;\;,\nl
J_{3,1}^{(k\ge3)} &=& {\La^2\over(k-1)(k-2)} - {2\mu\La\over k-2} + \mu^2L
- \mu^2\left(\wph\psi(k-2) + \geuler - {3\over2}\right)\;\;\;,\nl
J_{3,2}^{(k\ge2)} &=& {\La\over k-1} - 2\mu L +\mu\left(\wph\psi(2k)+\psi(2k-2) + 2\geuler\right)\;\;\;,\nl
J_{3,3}^{(k\ge1)} &=& L - \psi(3k) - \geuler - {3\over2}\;\;\;,\nl
J_{3,4}^{(k\ge0)} &=& {1\over3\mu}\;\;\;.
\eqa
We can draw the following conclusions. The convergent integrals do not depend on the
number of PV subtractions, as was to be expected. In the regularized divergent integrals
the only term that does not depend on the number of subtractions, and can be considered
unambiguous, is the logarithmic divergence $L$. Quadratic ($\La$) and higher ($\La^2,\ldots$)
divergences {\em do\/} depend on $k$ and are therefore ambiguous. Their coefficients 
approach zero with increasing number of subtractions. The finite terms are also ambiguous
(as was already remarked in \cite{Tim}),
and increase harmonically in absolute value with $k$. This is evidenced by the
ubiquitous $\log(1-x)/(1-x)$ in \eqn{Hvalues}. Strictly speaking, therefore, the limit
$k\to\infty$ is not clearly defined. We have checked that these features persist for
larger values of $\om$.\\

%\section{Resolving the ambiguities }
The reason why, in the previous section, we restricted ourselves to $\om\le3$ is that
the scalar $\f^4$ theory is renormalizeable for $\om=2$ and superrenormalizeable for
$\om=1$, and the $\f^3$ theory is renormalizeable for $\om=3$ and superrenormalizeable
for $\om=1,2$. We therefore consider the renormalization properties of the integrals $J$.
Obviously, we must apply a sufficient number of PV subtractions to properly regularize
these; but there is no obvious recipe for determining when to stop subtracting. The
only reasonable approach therefore seems to consider the case of an asymptotically
large number of subtractions, {\it i.e.\/} to take $k\to\infty$ in a sensible way. This has the
added advantage of being applicable to theories in {\em any\/} positive dimension. We can
use the fact that, asymptotically,
\bq
\psi(z) \approx \log(z) + \order{1/z}\;\;.
\eq 
so that $L-\psi(k)\approx\log(\La/\mu k)$.
In this limit we can write the regularized divergent integrals as
\bqa
J_{1,1} &=& \log\left({\La\over\mu k}\right) - \geuler\;\;\;,\nl
J_{2,1} &=& -\mu\log\left({\La\over\mu k}\right) +\mu\left(\wph\geuler-1\right)\;\;\;,\nl
J_{2,2} &=& \log\left({\La\over\mu k}\right) - \geuler -\log(2) - 1\;\;\;,\nl
J_{3,1} &=& \mu^2\log\left({\La\over\mu k}\right) - \mu^2\left(\geuler - {3\over2}\right)\;\;\;,\nl
J_{3,2} &=& -2\mu\log\left({\La\over\mu k}\right) + 2\mu\left(\wph\geuler+ \log(2)\right)\;\;\;,\nl
J_{3,3} &=& \log\left({\La\over\mu k}\right)- \geuler - \log(3)  - {3\over2}\;\;.
\eqa
Under renormalization, the logarithmic term is of course absorbed (and, in the spirit of
MS versus $\overline{\mbox{MS}}$, perhaps the $\geuler$ as well), and the results will
be well-defined and unambiguous.

A possible objection against the above procedure might be that 
the logarithmic divergence still contains $k$, albeit in
a more-or-less hidden manner. We might therefore choose to let $\Lambda$ depend on
the degree of PV subtraction as well, by writing 
\bq
\Lambda=\La_0k\;\;.
\eq
This choice contains a certain justice in that when we apply  more
PV subtractions the subtraction propagators individually become smaller.
In that case the higher divergences survive, but the results are still unambiguous:
\bqa
J_{1,1} &=& \log\left({\La_0\over\mu}\right) - \geuler\;\;\;,\nl
J_{2,1} &=& \La_0 - \mu\log\left({\La_0\over\mu}\right) + \mu\left(\wph\geuler - 1\right)\;\;\;,\nl
J_{2,2} &=&\log\left({\La_0\over\mu}\right) - \geuler -\log(2) - 1\;\;\;,\nl
J_{3,1} &=& 
{\La_0}^2 - 2\mu\La_0 + \mu^2\log\left({\La_0\over\mu}\right) - \mu^2\left(\geuler - {3\over2}\right)\;\;\;,\nl
J_{3,2} &=& \La_0 - 2\mu\log\left({\La_0\over\mu}\right) + 2\mu\left(\wph\geuler + \log(2)\right)\;\;\;,\nl
J_{3,3} &=& \log\left({\La_0\over\mu}\right)- \geuler - \log(3)  - {3\over2}\;\;.
\eqa
This approach works because the coefficients of $\La^n$ go as $k^{-n}$ for large $k$. We have
checked that this persists for larger values of $\om$.\\

As an example, we can apply the above strategy to {\it e.g.} the calculation of the
electron one-loop self-energy and the one-loop vertex correction in QED. In comparison
with the standard treatment as given in \cite{bibles} we find that our subtraction scheme results in
the following modification (in our notation):
\bq
\log(\La)\;\;\to\;\;\log(\La_0) - \geuler
\eq
in both computations. This shows that the relation between the vertex- and the wavefunction-
renormalization remains undisturbed; in particular the Ward-Takahashi identity remains valid for
the regularized Green's functions.\\ 

The authors are indebted to Prof. W. Beenakker for enlightening discussions.

\end{document}